\documentclass[pre,amsmath,amssymb,twocolumn,showpacs,nofootinbib]{revtex4-1}
\usepackage{graphicx,color}
\usepackage{etoolbox,hyperref}
\usepackage{dcolumn}
\usepackage{bm}
\usepackage[english]{babel}

\newcommand{\be}{\begin{equation}}
\newcommand{\ee}{\end{equation}}
\newcommand{\bea}{\begin{eqnarray}}
\newcommand{\eea}{\end{eqnarray}}

\newcommand{\nn}{\nonumber }

\begin{document}

\title{Statistics of shocks in a toy model with heavy tails}
\author{Thomas Gueudr{\'e} and Pierre Le Doussal} \affiliation{CNRS-Laboratoire
de Physique Th{\'e}orique de l'Ecole Normale Sup{\'e}rieure \\
24 rue Lhomond,75005 Paris, France} 
\date{\today\ -- \jobname}


\begin{abstract}
We study the energy minimization for a particle in a quadratic well in presence 
of short-ranged heavy-tailed disorder, as a toy model for an elastic manifold. The discrete model is shown to
be described in the scaling limit by a continuum Poisson process model which captures the three universality classes.
This model is solved in general, and we give, in the present case (Frechet class), detailed results for the 
distribution of the minimum energy and position, and the distribution of the sizes of the shocks (i.e. switches in the ground state)
which arise as the position of the well is varied. All these distributions are found to exhibit heavy tails with modified exponents. 
These results lead to an "exotic regime" in Burgers turbulence decaying 
from a heavy-tailed initial condition. 
\end{abstract}
\maketitle

\section{Introduction and model} 

Strongly pinned elastic objects, such as interfaces, occur in nature in presence of substrate impurity disorder which exhibits large fluctuations. The ground state
configuration is determined by a competition between the energy cost of deforming the interface and the energy gain in exploring larger regions of disorder. In the well studied case of Gaussian disorder, no impurity site particularly stands out and the optimum arises from a global optimisation. The typical interfaces are rough, with non-trivial roughness exponents $u \sim L^\zeta$, where $u$ is the deformation field and $L$ an internal coordinate scale. The optimal energy fluctuates from sample to sample with another exponent $H \sim L^\theta$. For directed lines (i.e. internal dimension $d=1$) wandering in one dimension, $\zeta=2/3$ and $\theta=1/3$, which in turn are related to the exponents of the standard universality class for the Kardar Parisi Zhang growth equation [\onlinecite{KPZ}]. 

In some physical systems however, the picture is completely different: a small fraction of the impurity sites produce a finite contribution to the total pinning energy, and the interface is deformed over large macroscopic scales, pinned specifically on those particular regions. One can see realisations of that situation in various area such as transition in chemical reaction of BZ type, or in granular flows [\onlinecite{atis_self-sustained_2012}]. One expects that the usual critical exponents are modified, but much less is known in this case, both about equilibrium (e.g. ground states) and about non-equilibrium dynamics (e.g. depinning). 

The present paper focuses on heavy-tailed disorder, which is paradigmatic of that situation, and whose probability distribution function \footnote{Also called probability density function below} (PDF), $P(V)$, shows an algebraic tail. In terms of the cumulative distribution function (CDF), denoted $P_{<}(V)= \int_{-\infty}^V P(V') dV'$ we have:
\begin{align} \label{tail} 
P_{<}(V)  \simeq \frac{A}{(-V)^{\mu}} \text{   for  } V \to - \infty
\end{align}
As was found in numerous works, such a scale-free distribution often leads to behaviors dominated by rare events. They have been much studied in the context of diffusion in random media, where they generate anomalous diffusion [\onlinecite{bouchaud1990anomalous}]. More recently heavy-tailed randomness was studied in the context of spin-glasses and random matrices [\onlinecite{burda_random_2006,janzen_levy_2010}]. For instance in [\onlinecite{biroli_top_2007}] it was found that the PDF of the maximal eigenvalue of a large random matrice with i.i.d entries distributed as  changes from the standard Tracy-Widom distribution (the Gaussian universality class) to a Frechet distribution as $\mu$ is decreased below $\mu=4$. 

Only a few works address the pinning problem in presence of heavy tails. In [\onlinecite{biroli_top_2007}] it was argued, based on a Flory argument, that for a directed polymer in the so-called $1+1$ dimensional geometry (meaning internal dimension $d=1$
and displacement $u \in R^D$ with $D=1$), for $\mu<5$ the roughness and energy exponents at $T=0$ change to $\zeta=(1+\mu)/(2 \mu-1)$ and $\theta=3/(2 \mu-1)$. For $\mu \geq 5$ one recovers the above mentioned values for Gaussian disorder, i.e. the tails have subdominant effect. While some mathematical results are available for $\mu<2$ [\onlinecite{hambly}], little is presently known rigorously for general $\mu$ or on the effect of a non-zero temperature on the problem [\onlinecite{auffinger_directed}]. 

In this paper we solve the much simpler case of a particle, which can be seen as the limit $d=0$ of the elastic interface problem. We consider the minimization problem:
\bea
&& H(r) = \min_u H(r,u) = H(r,u(r)) \\
&&  H(r,u) = \frac{m^2}{2}  (u-r)^2 + V(u)  \label{model1}
\eea 
where $V(u)$ is a random potential (a random function of $u$) and we define $u(r)={\rm argmin} H(r,u)$ the position of the minimum. The quadratic term confines the position $u$ of the particle and mimics the
elastic term for interfaces. More precisely, this model can be extended to an interface in a quadratic well and
there $m$ sets an internal length $L_m=1/m$ [\onlinecite{LeDoussalWiese2008c}]. Hence one can again define the exponents, as $m \to 0$:
\bea
u(r) - r \sim m^{-\zeta} \quad , \quad
H(r) - \overline{H(r)} \sim m^{- \theta} 
\eea 
where we denote by $\overline{\cdots}$ the average over the disorder. 
s
\begin{figure}[htb] 
\centering 
\def\svgwidth{240pt} 
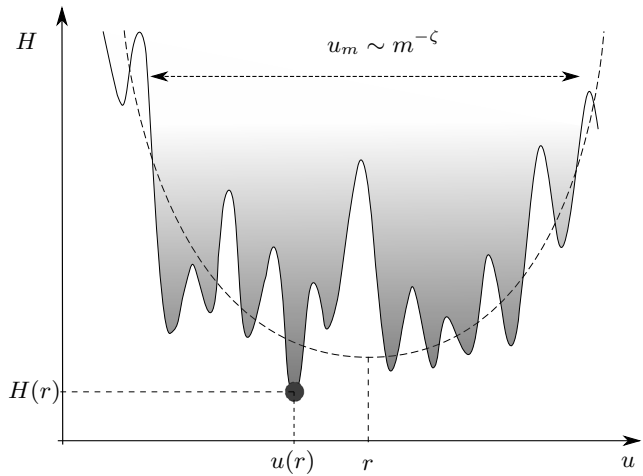
\caption{A particle in a random potential landscape confined by an elastic force (i.e. a quadratic potential centered at $r$). $u(r)$ is the position with
minimal total energy $H(r)$. Its fluctuations from sample to sample scale as $u_m \sim m^{-\zeta}$.}
\label{fpdf}
\end{figure}

This "toy model" has been much studied in the context of disordered systems for Gaussian disorder. It also
appears in the context of the decaying Burgers equation with random initial conditions, in the limit of vanishing viscosity
(see Appendix \ref{app:Burgers}  for details of the mapping). 
The case of short range correlations corresponding to a short-range potential $V(u)$ was solved in the seminal
paper of Kida [\onlinecite{Kida}]. An elegant derivation
using replica was also given in [\onlinecite{BM}]. Other derivations are given in [\onlinecite{LeDoussal2008}] (Appendix J)
and [\onlinecite{Bernardbauer}] (Appendix A). The case of Brownian correlations for $V(u)$ is related to the 
Sinai model studied in [\onlinecite{sinai_limiting_1983,LeDoussal2008,LeDoussalMonthus2003,burgers_non-linear_1974,frachebourg2000exact,valageas_statistical_whitenoise}]. Other type of correlations
have been studied in [\onlinecite{V,FyodorovLeDoussalRosso2010,Br,she_inviscid_1992,valageas_statistical_2009}].

Here we consider the case where:
(i) correlations of $V(u)$ are short range (ii) the PDF of $V(u)$ contains heavy tails. We then ask how the exponents and the PDF of $u(r)$ and $H(r)$ depend on the heavy-tail exponent $\mu$. Another interesting observables are the
jumps of the process $u(r)$. Indeed in the limit of small $m$ the process $u(r)$ consists mostly of jumps called
"static avalanches" or shocks (see below), and one define the shock sizes $s= u(r^+)-u(r^-)$.

To be specific we solve here two variants of the model:
\begin{itemize}
\item[(i)]
the discrete model: one starts with $u$ on a discrete lattice and i.i.d random variables $V(u)$. In the limit $m \to 0$ by rescaling the position $u$ the process converges to a continuum limit.
\item[(ii)]
the second is defined directly in the continuum for $u$: there $V(u)$ is defined as a Poisson point process.
\end{itemize}
Both models enjoy the same universal scaling limit.

In the absence of the quadratic well, $H = \min_u V(u)$ and the discrete problem reduces to the standard extreme value statistics problem.
It must then be defined for a fixed system size $u=1,..N$. For i.i.d random variable (or weakly correlated ones) 
$H$ then grows to infinity with the system size $N$ and, after a proper rescaling, the PDF of 
$a_N H + b_N$ converges to one of the famous three universality classes [\onlinecite{haan_extreme_2007}] : (i) Gumbel when 
$P(V)$ decays faster than a power law (ii) Frechet of index $\mu$ when $P(V)$ decays 
as a power law (\ref{tail}), and Weibull when $P(V)$ vanishes below some threshold (e.g. for $V<0$).
In the presence of the confining quadratic well, the same three classes survive: the Kida case belongs to the Gumbel class, 
while the heavy tail case belongs to the Frechet class. There are however some new universal features, such as the exponents and the distributions of shock sizes and minimum position. 

In this paper we derive a general formula for the PDF of the position of the minimum $u(r)$, and for the distribution of the shock sizes $s$. Although our formula is valid for the three universality classes, we give a detailed calculation in the case of the Frechet class with power law exponent $\mu$. 
We find that both distributions exhibit algebraic tails with modified exponents. These results are extended to space dimension $D>1$.

Note that some of our results were anticipated in the context of the decaying Burgers equation. 
In [\onlinecite{dbernard}] Bernard and Gawedzki looked for universality
classes distinct from Kida for statistically scale invariant velocity fields: they focused on the 
Weibul class and called it an "exotic regime"
for Burgers turbulence. In [\onlinecite{gurbatov_universality_2000}], a more general study was presented, encompassing
the three regimes. However, in none of these works the distribution of the shock sizes was obtained. 
The present work thus gives new results on another "exotic regime" for decaying Burgers
turbulence.

Note that the non-equilibrium version of this toy model, where one studies the dynamics of a particle
pulled quasi-statically by the harmonic well in the random potential $V(u)$ was studied in  [\onlinecite{LeDoussalWiese2008a}]. 
The three universality classes were also found to appear, and the distribution of the avalanche sizes were obtained
for the three classes. 

In Section \ref{discrete}, we solved the discrete toy model and obtain the joint PDF of the energy and the position in the small
$m$ limit. In Section \ref{univclass} we consider the Poisson process model, and derive the shock size distribution.
In Section \ref{highdimension}, we consider the discrete toy model in higher dimension. Finally, in Section \ref{floryargument}, we
discuss the case of a more general elastic manifold of internal dimension $d$ using
Flory arguments. The Appendix contains the mapping to Burgers, and mode details. 

\section{From the discrete model to the continuum: one point distributions} \label{discrete}

\subsection{Scaling exponents and dimensionless units} 

We now start from the discrete model where $u \in Z$ and $V(u)$ are i.i.d random variables drawn from
the distribution $P(V)$. We show that one obtains a non-trivial continuum limit in the limit $m \to 0$ upon rescaling of 
$u$ (in what we call dimensionless units below). This procedure makes the universality appear clearly.

Let us study first the one point distributions. For that purpose we can set $r=0$ and consider $H=H(r=0)$. The probability that the minimum total energy $H$ is attained in position $u$ with a value of the disorder $V$ is equal to the product of (i) the probability $P(V)$ of having $V$ in $u$ and (ii) the probability to have higher total energies on all the other sites $u'\neq u$. It is thus given by the infinite product: 
\begin{align} \label{infi_prod}
&p(u,V)=P(V) \prod _{u'\neq u} P_{>}(H-\frac{m^2 u'^2}{2})
\end{align} 

To study the limit of small $m$, it is convenient in the following to absorb the dependence with $m$ in the units $(u_m,H_m,V_m)$ defined for the variables $(u,H,V)$ respectively. One can then recover the dimensionful results by the substitution in all dimensionless results:
\begin{align}
u \rightarrow u/ u_m = m^{\zeta} u\\
V \rightarrow V/V_m = m^{\theta} V\\
H \rightarrow H/H_m = m^{\theta} H
\end{align}
Except if stated, we work now in the dimensionless system of units defined above. 
Without loss of generality, $A$ in Eq.\ref{tail} has been set to $1$ by a rescaling of $V$.

At this stage the exponents $\theta$ and $\zeta$ are not specified. To obtain a non-trivial limit one needs to scale
$V$ as $m^2 u^2$ which imposes the exponent relation:
\bea \label{sts} 
\theta = 2 \zeta -2 
\eea 
which is known in the directed polymer context as the STS relation [\onlinecite{hwafisher}].

The joint PDF Eq.\ref{infi_prod} for the optimal position $u$ and the value of the random potential $V$  on the optimal site 
then becomes, in the small $m$ limit:
\begin{align} \label{infi_prodnew}
& p(u,V)=m^{-\zeta-\theta} P(m^{-\theta} V) \prod _{u'\neq u}  P_{>}( m^{-\theta} (H - \frac{u'^2}{2}))\\
&\approx \frac{\mu}{|V|^{1+\mu}} \exp \left( - \int du' m^{-\zeta} P _< ( m^{-\theta} (H -\frac{u'^2}{2}))\right) \theta_{H<0} \\
&\approx \frac{\mu}{|V|^{1+\mu}}\exp \left( - F_\mu |V+\frac{u^2}{2}|^{\frac{1}{2} - \mu}\right)  \theta_{V+\frac{u^2}{2}<0} 
\end{align}
where $H=V+ \frac{u^2}{2}$ and we denote everywhere $\theta_{x<0}$ the characteristic function of the interval (Heaviside function).
Here and below we denote:
\be \label{Fmu} 
F_\mu=\frac{\sqrt{2 \pi} \Gamma [\mu - 1/2]}{\Gamma [\mu]}
\ee
The joint PDF of $u$ and $H$ is simply $p(u,V=H- \frac{u^2}{2})$. Going from the infinite product to the exponential in the second line of Eq.\ref{infi_prodnew} requires that $P_{>}(\cdot) \sim 1$ at all
sites, or equivalently $H< 0$, which is verified for $m$ small enough.
The final expression for the joint PDF Eq.\ref{infi_prodnew} is normalized to unity $\int dV du p(u,V)=1$, which shows that
we have correctly taken the small mass limit (no regions have been overlooked). 
More precisely, and as is further explained in the Appendix \ref{app:infinite}, as $m \rightarrow 0$ (the continuum limit), the rescaled cumulative (CDF) $m^{-\zeta} P _< ( m^{-\theta} y)$ converges to $\frac{\theta_{-y}}{(-y)^{1+\mu}}$ (under the condition that the right tail is in $o(V^{-(1+\mu)})$, cf.
Appendix \ref{app:infinite}). 
Hence only the contribution of the left tail of $P_<(\cdot)$ contributes to the integral in Eq.\ref{infi_prodnew}, a typical behaviour in power law statistics and on can readily replace $P_<(\cdot)$ by its asymptotic expression (such estimates can be established rigorously by the use of tauberian theorems [\onlinecite{daleyhall}]). This implies the second relation:
\bea
\zeta = \mu \theta 
\eea 
which leads to:
\bea  \label{exponents} 
&& \zeta = \frac{2 \mu}{2\mu - 1} \\
&& \theta = \frac{2}{2\mu - 1}
\eea
One notes that, unlike the directed polymer, there is no finite critical value of $\mu$ at
which one recovers the Gaussian behavior. In other words any power law tail matters.
More precisely one can say that $\mu_c=+\infty$. In that limit, indeed, $\zeta \to 1$
which is the value for the Gumbel class [\onlinecite{LeDoussal2008}]. There is
an interesting crossover in that limit where the leading contribution goes from
the bulk of $P(V)$ (as is the case for the Gumbel class) to the tail (for the present
power law case).

\subsection{Results for the one-point distributions}

From Eq.\ref{infi_prod}, one can obtain the joint distribution of $(H,V)$. Taking into account the jacobian $\frac{\partial (u,V)}{\partial(H,V)}= (\sqrt{2} (H-V))^{-1/2}$ and a factor of $2$ from integration over positive and negative $u$ yields:
\begin{align}
p(H,V)=\frac{\mu \sqrt{2} }{|V|^{1+\mu} \sqrt{H-V}} e^{ -F_\mu |H|^{\frac{1}{2} - \mu}} \theta_{H<0,V<H} 
\end{align}
After integration, one obtains the various marginal distributions of $H$, $V$ and $u$. First we
obtain:
\bea
p(H) = \frac{(\mu - \frac{1}{2}) F_\mu}{|H|^{\mu + \frac{1}{2}}} e^{-F_\mu |H|^{\frac{1}{2}-\mu}} \theta_{H<0} 
\eea 
Hence, the PDF of the total energy $H$ is a Frechet distribution. On one hand this appears as natural since
we are dealing with extreme value statistics of heavy tailed distributions. However, the index of the Frechet
distribution is not $\mu$ (as would be naively expected) but $\mu -1/2$, which is thus a correction coming from the competition
with the elastic energy. As the particle chooses amongst the deepest sites, the distribution of its energy 
acquires a power-law tail which is even broader than the initial disorder. It is easy to extend the
above calculation to a generalized elastic energy growing as $u^{\alpha}$, the modified index being then 
$\mu - 1/\alpha$. 

Next we also obtain the PDF of the potential $V$ at the position of the minimum as:
\begin{align}
p(V)&= \frac{\mu  }{|V|^{\mu +1}} \phi _{\mu} (|V|) \theta_{V<0} 
\end{align}
where we have defined the auxiliary function:
\begin{align}
\phi _{\mu} (x)&= \sqrt{2} \int _0 ^x \frac{dy}{\sqrt{x-y}} e^{-F_\mu y^{\frac{1}{2}-\mu}} 
\end{align} 
Note that the factor $\phi_\mu(|V|)$ gives the relative change of the tail of the PDF of the potential at the optimal site w.r.t.
the tail of the original PDF of the disorder. For $|V|$ of order one it is of order one, hence the original tail
exponent is not changed but the amplitude is changed \footnote{One should keep in mind that here $V$ denotes the dimensionless
potential hence it is deep in the tail since we use units of $V_m \sim m^{-\theta}$.}. For large negative $V$ it diverges hence we find:
\bea \label{pV}
p(V) \simeq \frac{2 \sqrt{2} \mu}{|V|^{\mu+ \frac{1}{2}}} \quad , \quad V \to - \infty
\eea 
which is again the original tail but with the same shift in the exponent $\mu \to \mu-\frac{1}{2}$ as
noticed above, and a different amplitude.

\begin{figure}
\includegraphics[scale=0.25]{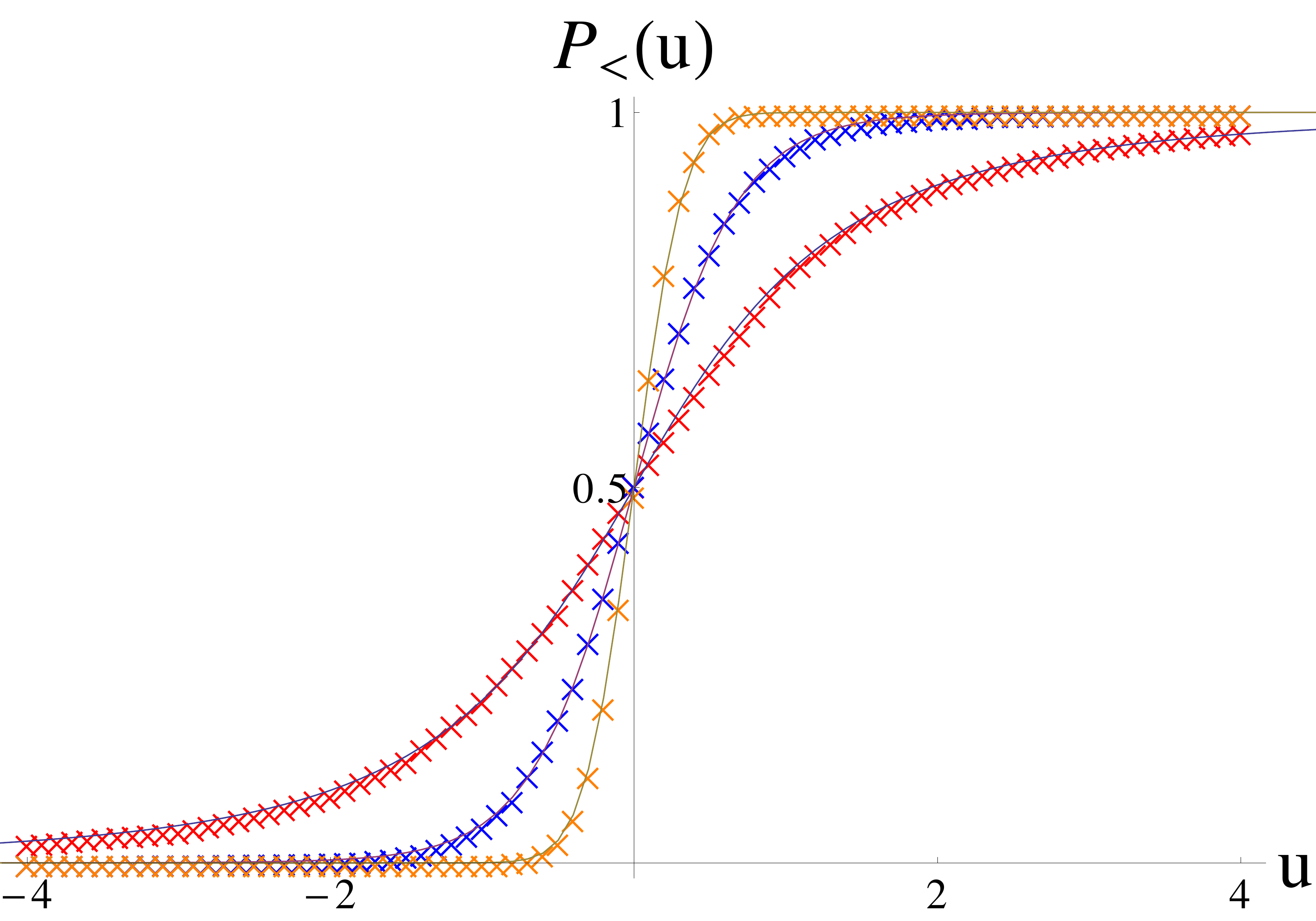}
\centering
\caption{(Color Online). Comparison of the  CDF for the position $u$ (full line) with numerical simulations (crosses). From the steepest curve to the least steep, $\mu = 8 \text{ (orange), } 4 \text{ (blue), } 1.5 \text{ (red) } $. The sample size is $N=10 ^6$.}
\label{pospdf}
\end{figure}

Finally we obtain the PDF of the optimal position $u$ of the particle as:
\begin{align}  \label{Pu}
p(u)&= \mu \ \psi_{\mu} (\frac{u^2}{2})
\end{align}
in terms of the auxiliary distribution:
\begin{align} \label{psi}
\psi _{\mu}(x)&= \int _0 ^\infty \frac{dy}{(x+y)^{\mu +1}} e^{-F_\mu y^{\frac{1}{2} -\mu}}
\end{align} 
The PDF of $u$ decreases from a constant at $u=0$ to a power law at large $u$. 
The position of the particle is thus heavy tailed as well as its PDF decays as 
\bea
p(u) \simeq \frac{2^{\mu}}{u^{2 \mu}}  \quad , \quad |u| \to + \infty
\eea
The moments $\overline{u^{2n}}$ thus exist only for $2 n < 2 \mu-1$ and are
given in the Appendix \ref{app:moments}. The comparison with numerics is made on Fig.\ref{pospdf}.
Finally note that for $\mu<\frac{1}{2}$ the particle explores the whole space 
$u \sim W$, as the energy of the optimal site $\sim u^{1/\mu}$ grows faster than the elastic energy
$\sim u^2$. 

We note that the PDF of the "elastic energy" $E=u^2/2$ has also a tail:
\bea
p(E) \simeq \frac{1}{\sqrt{2}} \frac{1}{E^{\frac{1}{2} + \mu}}
\eea
with exponent $\mu-\frac{1}{2}$
analogous to (\ref{pV}) for large values.

To conclude, the typical $H,V$ of order one are already
drawn in the original tail of $P(V)$ with exponent $\mu$ (since we work in the units $m^{-\theta}$) 
and the rare events acquire a tail with exponent $\mu-\frac{1}{2}$.

\section{Statistics of the shocks} \label{univclass}

As the center of the harmonic potential $r$ is shifted, the optimal position $u(r)$ of the particle is changed as shown
in Fig.\ref{shockpic}. This corresponds to a jumpy motion of the particle, each jump is called a shock because corresponding to traveling shocks in the Burgers velocity field (see Appendix \ref{app:Burgers}). We now introduce the Poisson process model. 

\subsection{The general case}

\subsubsection{Poisson process model and one-point distribution} 

The computation on the discrete model being rather cumbersome, we follow [\onlinecite{dbernard}] and start directly in the continuum by distributing the random energies over the line as a Poisson process over the plane $(V,u)$ of density $f(V)dVdu$. Each cell of size $dV du$ is then either occupied or not, depending on the value of the random potential $V_i$ at site $u_i$. This means that the potential is defined only at the $u_i$ with values $V(u_i)=V_i$ and that:
\begin{align}
\label{minHj}
H(r) &= \min_j H_j(r) = \min_j \left(V_j +  \frac{(u_j-r)^2}{2} \right)\\
u(r) &= {\rm argmin } ~ H_j(r)  \label{minj} 
\end{align}
We denote the primitive $F(x)=\int_{- \infty} ^x f(t) dt $ and assume that
$F(+\infty)=+\infty$. 
We now calculate, using methods similar to the one of [\onlinecite{dbernard}], the 
one and two point characteristic function of the field $u(r)$. 

For the one point function we can choose $r=0$, and define $u=u(0)$. 
Using formulas similar to Eq.\ref{infi_prod} we find for the
joint distribution of position and potential at the minimum:
\begin{align}
&p(u,V) dV du = f(V) dV du \\
&\prod _{d V' du'} \left( 1 - \theta_{V'+\frac{u'^2}{2} < V + \frac{u^2}{2}} f(V')dV' du' \right) \nn
\end{align}

From the infinitesimal version of Eq.\ref{infi_prod}, and after the change of variables
$z= u' \text{, }  \phi = V + \frac{u^2}{2}$, the one-point distribution of the 
position of the minimum can be expressed as:
\begin{align}
\label{positiondensity}
p(u) = \int d \phi  f (\phi - \frac{u^2}{2}) \exp \left( - \int  dz F(\phi - \frac{z^2}{2}) \right) 
\end{align}
It is easy to check the normalization $\int du p(u) =1$ by noting that the integral is a total derivative. 
This result is valid for arbitrary Poisson measure $f(V)$. As we discuss below
one can recover the results of the previous section in a particular case. 

\begin{figure}[htb] 
\centering 
\def\svgwidth{240pt} 
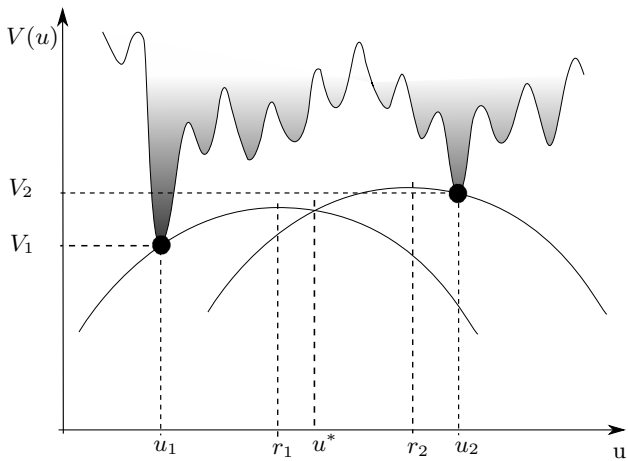
\caption{The parabola construction for the minimisation problem: when the center $r$ of the parabola is shifted from $r_1$ to $r_2$, the position of the particle moves from $u_1$ to $u_2$. For given $r_1$ and $r_2$, the intersection of both the parabola is called $u^*$.}
\label{shockpic}
\end{figure}

\begin{figure}[htb] 
\centering 
\def\svgwidth{240pt} 
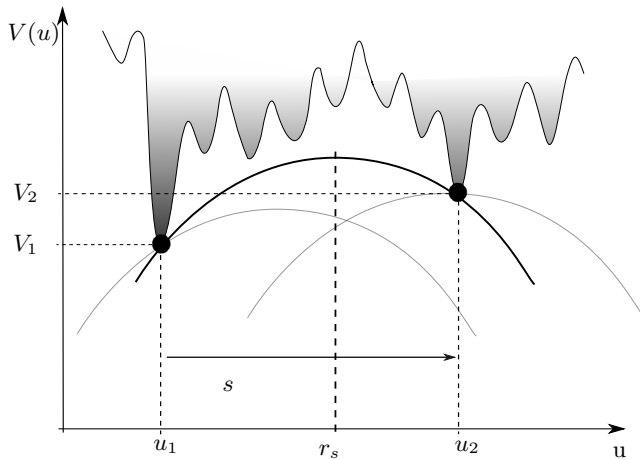
\caption{The discontinuous motion of the particle can be decomposed in shocks. Those shocks occur (here in $r_s$) while the parabola is shifted and touches the potential at two positions $u_1$ and $u_2$, as depicted. The size of the shock is denoted $s= u_2 - u_1$.}
\label{shockpic3}
\end{figure}

\subsubsection{Shock and droplet size distributions} 

To describe the statistical properties of the jumps of the optimal position $u(r)$ of the particle as $r$ is varied one
defines the shock density as:
\bea
\rho(s) = \lim_{\delta r \to 0^+} \frac{1}{\delta r} \overline{\delta(u(r+\delta r) - u(r) - s) } 
\eea 
Another definition, equivalent in the present case, uses the decomposition:
\bea
u(r) = \sum_i s_i \theta_{r>r_i} + \tilde u(r)
\eea 
where $\tilde u(r)$ is the smooth part of the field $u(r)$, which, for the Poisson process model can be set to zero. For other models
in the same universality class this part is subdominant. The 
shock density is then defined as [\onlinecite{LeDoussalWiese2008c}]:
\bea
\rho(s) = \overline{ \delta(r-r_i) \delta(s-s_i)} 
\eea 
where the $(r_i,s_i)$ are the positions and sizes of the shocks. Note that all the $s_i>0$. 

The shock density is intimately related to another quantity, the droplet density $D(s)$, namely the probability density for the total energy 
$H_j(r)$ in (\ref{minHj}) for a given $r$, to exhibit two degenerate minima at positions $u_1$ and $u_2$, separated in space by $s=u_2-u_1$ (see Fig.\ref{shockpic3}).
By construction $D(s)$ is a symmetric function $D(s)=D(-s)$ and has dimension $1/(s E)$ where $E$ is an energy. More precisely, it is defined as 
$D(s)=\int du_1 du_2 \delta(s-u_2+u_1) p(u_1,u_2,0)$ where $p(u_1,u_2,E)$, the is probability density for the absolute minimum in $u_1$ and the secondary minimum in $u_2$ separated by $E>0$ in energy. The knowledge of this function allows to calculate all thermal
cumulants at low temperature (see e.g. [\onlinecite{MonthusLeDoussal2004,LeDoussal2008}]). 

As before, we denote the minimal total energy $\phi = H(u_1)=H(u_2)$. Requiring all the other sites to have higher total energy induces a factor $\exp \left(- \int F(\phi - z^2/2)dz \right)$ similarly to (\ref{positiondensity}). Then the integrated probability over the value $\phi$ of the minimum and the positions $u_1$ and $u_2$ at fixed $s= u_2-u_1$ lead to:
\begin{align}
\nonumber
D(s)= &\int d\phi du_1 du_2 f(\phi - \frac{u_1^2}{2})f(\phi - \frac{u_2^2}{2})\\ 
 &\exp \left( -\int F(\phi - \frac{z^2}{2})dz  \right) \delta (s-  u_2 + u_1)
\end{align}

The relation between the shock and the droplet density can be written 
(see Ref. [\onlinecite{LeDoussal2008}], Sections IV B.5 and E.4) for $s>0$:

\begin{align}
\label{shockdroplet}
\rho (s) = s D(s) \theta_{s>0} \\
\nonumber
\end{align}
The factor $s$ originates from the change of variable from energy to position as $\frac{\partial H}{\partial r}$ noting that
a small change in the position of the parabola around the point of degeneracy amounts to shift the
relative energies of the two states by:
\bea \label{shift} 
\delta H = \delta r \times (u_1-u_2) 
\eea 

Using this relation (\ref{shockdroplet}) we now obtain the shock density, which can be rewritten as, for $s>0$:
\begin{align}
\label{shock_pdf}
\nonumber
\rho(s) &= \frac{s}{2} \int d \phi \ dz f\left(\phi-\frac{(z-s)^2}{8} \right) \\
&f\left(\phi-\frac{(z+s)^2}{8}\right)e^{- \int dz' F(\phi -\frac{z'^2}{2})}
\end{align}
where we denoted $z=u_1+u_2$. 

From the shock density one can define a normalized size probability distribution as:
\bea
\rho(s) = \rho_0 p(s) 
\eea 
where $\int_0^\infty ds p(s) = 1$ and $\rho_0$ is the total shock density. The density
$\rho(s)$ satisfies the following "normalization" identity:
\begin{align} \label{norm1} 
\int_0^\infty ds ~ s \rho (s)  = 1
\end{align}
which expresses that all the motion occurs in the shocks. Similarly $D(s)$ satisfies $\int_{-\infty}^{+\infty} ds ~ s^2 D(s) = 2$. 
This identity, proved in the Appendix \ref{app:norma} is a signature of the STS relations which originate from the statistical translational invariance of the problem.

As a consistency check, $\rho(s)$ can also be extracted from the small separation behavior of the two point characteristic function of the position
field $u(r)$, for $r>0$:
\bea 
\nonumber
\overline{ e^{\lambda (u(r)-u(0))}} = 1 + r \int_0^\infty ds \rho(s) (e^{\lambda s}-1) + O(r^2)  \\
\label{id1} 
\eea 
The calculation of this function is more cumbersome and displayed in Appendix \ref{2pointfunction}.
As shown there, by identification in the above formula one recovers Eq.\ref{shock_pdf}.

\subsection{Scale invariance and universality classes}

From Eq.\ref{shock_pdf}, one can read the distribution of the shock sizes for any disorder in the continuum Poisson process model. 
For this model to be a "fixed point" (i.e. continuum limit) of a more general class of models (e.g. the discrete model studied in Section as $m \to 0$) one should in addition require scale invariance. Then, similarly to the usual problem of extremal statistics [\onlinecite{schehr_exact_2013}], and to the problem of the driven particle [\onlinecite{LeDoussalWiese2008a}], three different classes of universality emerge. 
The nice feature of the Poisson process model is that it contains the three scale invariant models. 

\subsubsection{The three universality classes}

Let us consider again the minimization problem (\ref{minj}) in a dimension-full form:
\begin{align}
H_m(r) &= \min_j (V_j +  m^2 \frac{(u_j-r)^2}{2} ) \label{minj2} 
\end{align}
Let us require that $H_m(r)$ is scale invariant {\it in law}, i.e. that $H_{m}(m^{-\zeta} r)$ has the same distribution
as $m^{-\theta} H_{m=1}(r)$, possibly up to an additive constant in $H$. One easily sees that it implies 
that $f(m^\theta V) = m^{-(\theta+\zeta)} f(V+C_m)$ and the STS exponent relation (\ref{sts}).
There are three type of solutions. 

\begin{itemize}
\item The "Gumbel" class, where the disorder left tail is exponentially fast decaying. This case corresponds to the well-known Kida statistics of the Burgers equation [\onlinecite{Kida}], and is obtained for a Poisson density $f(\phi) = e^{\phi}$ with the density of shocks:
\bea
\rho (s) = \frac{1}{2\sqrt{\pi}} ~ s ~ e^{-s^2/4}
\eea

\item The "Weibull" class, where the disorder is bounded from below. It corresponds to the Poisson process model with $f(\phi) = \frac{1}{\phi^{1+\mu}} \theta_{\phi>0}$ with $-\infty< \mu < -1$. This model was studied in [\onlinecite{dbernard}]. 

\item The "Frechet" class, the focus of the present paper, where the disorder presents an algebraic left tail, accounting for rare but large events. 
It corresponds to the choice $f(\phi)=\frac{1}{|\phi|^{1+\mu}} \theta_{\phi<0}$. As discussed above, this choice represents the continuous limit of the system defined in Section \ref{discrete}.

\end{itemize}

Note that in all three classes the exponents are given by (\ref{exponents}), the Gumbel class corresponding to $\mu=+\infty$ (with additional
logarithmic corrections in that case). 

We now study in more details the distribution of shock sizes in the Frechet class, and compare to the classical Kida statistics. 

\subsubsection{Shock size distribution in the Frechet universality class}

Let us consider the Poisson process model with the choice:
\begin{align}
f(\phi)&=\frac{\mu}{(-\phi)^{1+\mu}} \theta_{\phi<0} \\
\nonumber
F(\phi)&= \frac{1}{(-\phi)^{\mu}}  \theta_{\phi<0}  +  \infty  \times \theta_{\phi>0} 
\end{align}
With this choice one sees that the formula (\ref{positiondensity}) for $p(u)$ for the
Poisson model becomes identical (identifying $y=-\phi$) to formula (\ref{Pu}),(\ref{psi}) for the discrete model
with the same constant $F_\mu$ given by (\ref{Fmu}). Note that the exponential factor in (\ref{positiondensity}) vanishes 
if $\phi>0$ hence the $\phi$ integration is in effect restricted to $\phi<0$. 

We now consider the shock size distribution from (\ref{shock_pdf}):

\begin{align}  \label{rhov}
\nonumber
\rho(s) & = \mu^2 s \int_0^\infty dz  \int_{-\infty}^0 d\phi \exp \left( - F_\mu |\phi|^{\frac{1}{2}-\mu} \right) \\
&\times \left[\left(\frac{(z+s)^2}{8}-\phi \right) \left(\frac{(z-s)^2}{8}-\phi \right)\right] ^{-(1+\mu)}
\end{align}
and we assume here $\mu>1/2$. 

This function does not exhibit any divergence for small shock sizes, rather it behaves similarly to the Kida
distribution at small $s$ with:
\bea
\label{asymptsmall}
\rho(s) \simeq C_\mu s 
\eea 
and the constant $C_\mu$ is displayed in the Appendix \ref{app:asympt}. The main difference arises in the
behavior of the large shocks. Instead of the exponential tail $e^{-s^2/2}$ in the Kida case, it shows algebraic tails
of the form:
\bea
\label{asymptlarge}
\rho(s) \simeq \frac{2^{2+\mu} \mu }{s^{\tau'}} \text{  for large $s$}
\eea 
with the decay exponent $\tau'$ for the right tail \footnote{We use the notation $\tau'$ to
distinguish from the exponent for the divergence of {\it small} shocks usually called $\tau$.}:
\bea
\tau' = 1+ 2 \mu
\eea 
To obtain this result from (\ref{rhov})
one notes that it is the region for $z$ near $s$ which contributes most, hence one
shifts $z \to z+s$ in (\ref{rhov}) and replaces $\frac{1}{8}(z+2s)^2-\phi \to s^2/2$ in the first factor.
The remaining integral, can be extended from $z \in [-\infty,\infty]$ and can then be performed exactly, being related to the normalization of the
distribution $p(u)$ of a single minimum (\ref{Pu}): one uses $\int dz \psi_\mu(z^2/8)=2/\mu$. 
Note that since we assumed $\mu>1/2$ it implies that $\tau'>2$, hence the integral (\ref{norm1}) exists, as required. 
However the second moment of the shock size, $\int_0^{+\infty} ds s^2 \rho(s)$ is finite only for $\mu>1$ \footnote{In the
functional RG this quantity equals $- \Delta'(0^+)/m^4$, while the second moment of $p(u)$ in Eq. (\ref{Pu}) is $m^2 \overline{u^2} = \Delta(0)$
(which exists only for $\mu>3/2$) 
where $\Delta(u)$ is the correlator of the renormalized disorder (see \cite{LeDoussalWiese2008c,LeDoussal2008} for definitions).}.

\begin{figure}
\includegraphics[scale=0.33]{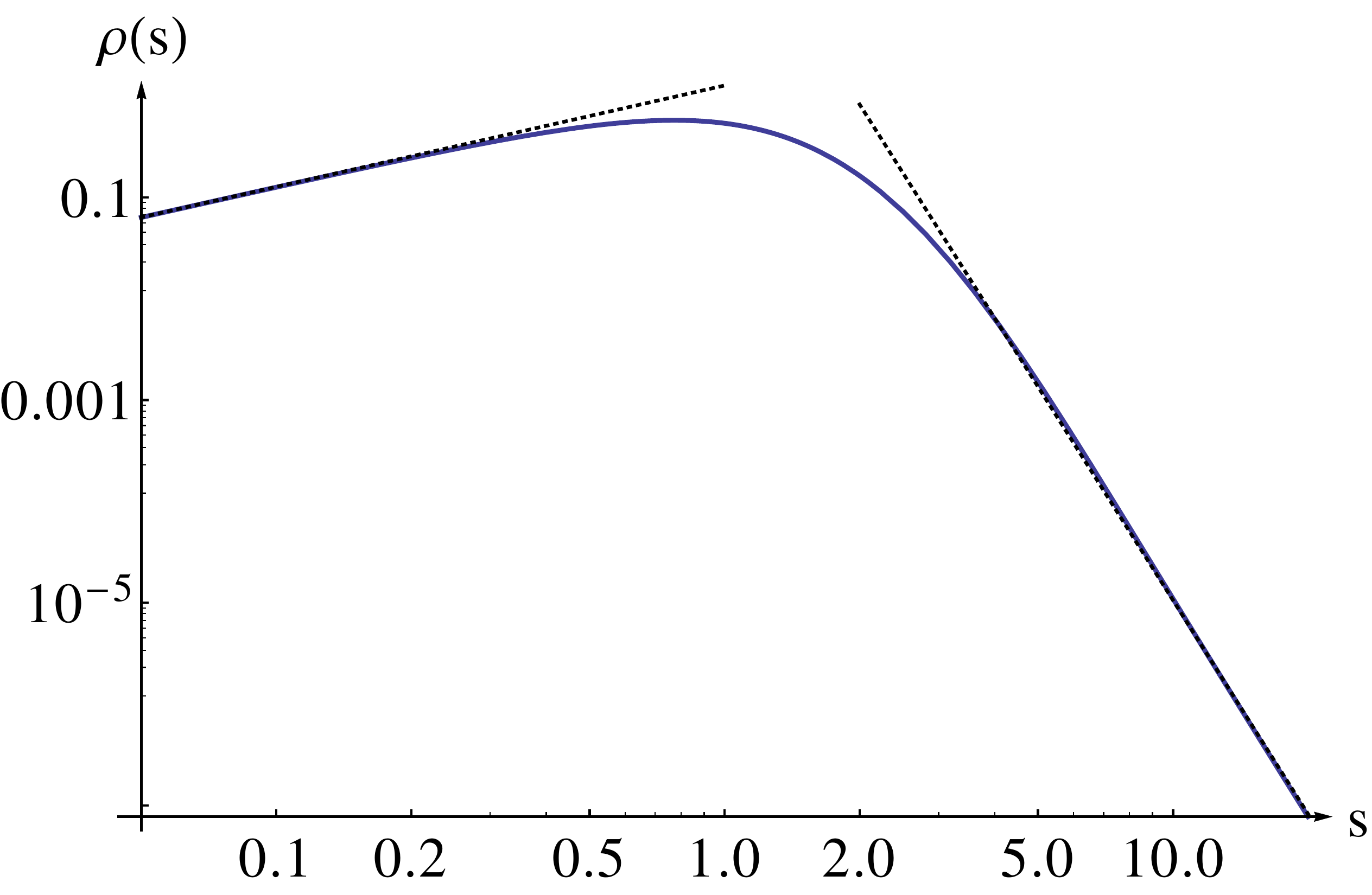}
\centering
\caption{The PDF $\rho (s)$ of the shocks size, plotted from Eq.(\ref{rhov}) for $\mu=3/2$. In black dotted lines are the asymptotics for small and large $s$ as given by Eqs.(\ref{asymptsmall}) and (\ref{asymptlarge}).}
\label{shockpdf}
\end{figure}

Finally it is useful to recall for comparison the avalanche size distribution for the non-equilibrium 
version of this model, i.e. the quasi-static depinning. There the jumps occur between the
metastable states actually encountered in the driven dynamics as $r$ increases, which are different from the
absolute energy minima. The result of  [\onlinecite{LeDoussalWiese2008a}] for the Frechet class 
for the normalized distribution is:
\bea
p(s) = \frac{(\alpha+1)(\alpha+2)}{\Gamma(2+\frac{1}{\alpha})} \int_0^{+\infty} \frac{dy}{(y+s)^{3+\alpha}} e^{- y^{-\alpha}} 
\eea 
where the local disorder {\it force} is short range distributed with a heavy tail index $\mu=1+\alpha$. The large
$s$ behavior is also a power law $p(s) \sim s^{-(2+\alpha)}\sim s^{-(1+\mu)}$.

\section{The model in dimension $D > 1$} 
\label{highdimension}

The methods of solution presented in the previous sections can be extended to the toy model of the particle (i.e. $d=0$) 
in general (external) space dimension ${\bf u} \in R^D$. The position of the minimum when the quadratic well is centered in
${\bf r} \in R^D$ is now denoted
as ${\bf u}({\bf r})$, a vector process which exhibits jumps, in fact it is constant on cells in $R^D$, separated by shock walls
with discontinuities where it jumps by ${\bf s}$. To generalize most of the calculations one must simply 
replace the integrals over the spatial variable $u$ by integrals over vectors $\textbf{u}$. 
The new scaling exponent necessary to retain invariance of the tail of the potential are:
\begin{align} \label{exponentsD} 
\zeta = \frac{2 \mu}{
  2 \mu-D}\\
\theta= \frac{2 D}{2 \mu -D} \\
\end{align}
which reduce to (\ref{exponents}) for $D=1$ and still satisfy the relation (\ref{sts}). 
Let us first discuss one point probabilities, hence setting ${\bf r}=0$. 

\subsection{One point distribution} 

Due to the rotational invariance of the elastic energy, one readily obtains
the joint distribution:
\bea
p(\textbf{u},V)=\frac{\mu}{|V|^{1+\mu}}e^{-F_{\mu,D} |H| ^{\frac{D}{2} - \mu}} \theta_{H<0} 
\eea 
where $H=V + \frac{u^2}{2}$. It is normalized to unity $\int d^D \textbf{u} d V p(\textbf{u},V)=1$ and we have
defined:
\bea
F_{\mu,D}=S_D 2^{D/2-1} \frac{\Gamma [D/2]\Gamma [\mu - D/2]}{\Gamma [\mu]}
\eea 
where $S_D$ is the surface of the unit sphere in dimension $D$ ($S^1 = 2$). 
From this we extract the joint distribution of $V$ and $H$ as:
\begin{align}
p&(V,H)= S_D 2^{\frac{D}{2}-1} (H-V)^{\frac{D}{2}-1} \frac{\mu}{|V|^{1+\mu}} \\
&\exp \left(- F_{\mu,D} |H|^{\frac{D}{2} - \mu} \right) \theta_{H<0,V<H}
\end{align}
which exhibit a "level repulsion" between $H$ and $V$ for $D>2$.

The marginal distribution for $H$ is again a Frechet with index now $\mu - \frac{D}{2}$:
\bea
p(H) = \frac{(\mu - \frac{D}{2}) F_{\mu,D} }{|H|^{\mu - \frac{D}{2}+1}} e^{-F_{\mu,D} |H|^{ \frac{D}{2} -\mu}} \theta_{H<0} 
\eea 
while the PDF of $V$ takes the form:
\bea 
p(V)= \frac{\mu S_D}{2^{1-D/2} |V|^{\mu +1}} \phi^D _{\mu} (|V|) \theta_{V<0}
\eea
where we have defined:
\begin{align}
\phi^D _{\mu} (x)&= \int _0 ^x \frac{dy}{(x-y)^{1-D/2}} e^{-F_{\mu,D} y^{\frac{D}{2} - \mu}} 
\end{align} 

Finally the distribution of the optimal position is:
\begin{align}
p(\textbf{u})&= \mu \ \psi^D_{\mu} (\frac{u^2}{2})
\end{align}
where:
\begin{align}
\psi^D _{\mu}(x)&= \int _0 ^\infty \frac{e^{-F_{\mu,D} y^{\frac{D}{2} - \mu}}}{(x+y)^{\mu +1}}
\end{align} 
and, interestingly the tail exponent of $P(\textbf{u})$ is independent of $D$:
\bea
p(\textbf{u}) \simeq \frac{2^{\mu}}{u^{2 \mu}}  \quad , \quad |u| \to + \infty
\eea
while the PDF for the radius $|u|$ decays as $\simeq 2^{\mu} S_D/|u|^{2 \mu+1 - D}$. 

Note that the condition for the thermodynamic limit to be defined is now $\mu > \frac{D}{2}$, as the typical minimum
site energy at a distance $u$ of the center grows as $u^{D/\mu}$. 

\subsection{Droplet and shock densities} 

Note that the formula for the droplet density also generalizes easily in $D$ dimension as:
\begin{align}
\label{highdroplet}
\nonumber
D({\bf s})= &\int d\phi d^D {\bf u_1} d^D {\bf  u_2} f(\phi - \frac{u_1^2}{2})f(\phi - \frac{u_2^2}{2})\\ 
 &\exp \left( -\int F(\phi - \frac{z^2}{2})d {\bf z}  \right) \delta^D ({\bf s}-  {\bf u}_2 + {\bf u}_1)
\end{align}
where $\vec s$ is the vector joining the two degenerate minima. It is now normalized as:
\bea
\int d^D {\bf s} ~ s^2 D( {\bf s}) = 2 D 
\eea
as shown in the Appendix \ref{app:norma}. The shock density is now defined by reference to a direction
of unit vector ${\bf e}_x$ as:
\bea
\rho({\bf s}) = \lim_{\delta r \to 0^+} \frac{1}{\delta r} \overline{\delta^D({\bf u}({\bf r}+\delta r  ~{\bf e_x}) - {\bf u}({\bf r}) - {\bf s}) } 
\eea 
Since (\ref{shift}) generalizes to $\delta H = \delta r ~{\bf e_x} \cdot ({\bf u}_1-{\bf u}_2)$ one sees that the relation between the shock and droplet densities is now:
\bea
\label{highshock}
\rho({\bf s}) = s_x D({\bf s}) \theta_{s_x>0} 
\eea 
where $s_x={\bf s} \cdot {\bf e}_x$ denotes the component of the jump along the direction $x$. 

Using isotropy it now enjoys the normalization:
\bea \label{normD} 
\int_{s_x>0} d^D {\bf s} ~ s_x \rho({\bf s}) = 1
\eea 
which, again, expresses that all motion when ${\bf r}$ varies along a line, occurs in shocks. Note that
the relation (\ref{highshock}), combined with the isotropy of $D({\bf s})$ implies a number of relations\footnote{These are easily shown e.g. by integrating w.r.t. $D({\bf s}) \to e^{- \mu s^2}$ since any isotropic distribution can be represented
as a superposition of such weights.} between
moments, for instance:
\bea \label{rel}
\langle s_x ^2 \rangle = 2 \langle s_y ^2 \rangle 
\eea
as well as $\langle s_x ^4 \rangle = \frac{8}{3} \langle s_y ^4 \rangle = 4 \langle s_x^2 s_y ^2 \rangle$ and so on {\it provided these
moments exist}, i.e. that the tail of $D({\bf s})$ decays fast enough\footnote{The relation (\ref{rel}) is believed to be more general
(i.e. to extend to interfaces) and was anticipated in [\onlinecite{PDLinprep}] where it was related via the functional RG to the existence of a cusp in the effective action of the theory
(see also [\onlinecite{LeDoussalRossoWiese2011}]).}. 

It is interesting to note that Eqs.(\ref{highdroplet}) and (\ref{highshock}) factorize in the Kida (i.e. Gumbel) universality class (i.e. with the choice $f(\phi)=e^{\phi}$) leading to
the simple result, after some Gaussian integrations:
\begin{align} \label{KidaD} 
\rho({\bf s})= \frac{s_x}{(4 \pi)^{\frac{D}{2}}} e^{-s_x^2/4}  e^{-{\bf s}_{\bot} ^2/4} 
\end{align}
where we denote ${\bf s} = (s_x,{\bf s}_{\bot})$ and $s_{\bot}$ represents the "wandering" part of the shock motion, 
transverse to the shift direction of the parabola. For instance in two dimension ${\bf s} = (s_x,s_y)$, Eq. (\ref{KidaD}) reads
$\rho({\bf s})=\rho _{D=1} (s_x) D_{D=1}(s_y)$. Hence in the Kida case, higher dimensions statistics of the shocks are completely solved from the $D=1$ case.

The Frechet case, however does not simplify as nicely. One now obtains:
\begin{align}  \label{rhov2}
\rho(s) & = \mu^2 \frac{s_x}{2^D} \int_0^\infty d^D {\bf z}  \int_{-\infty}^0 d\phi \exp \left( - F_{\mu,D} |\phi|^{\frac{D}{2}-\mu} \right) \\
&\times \left[\left(\frac{({\bf z}+{\bf s})^2}{8}-\phi \right) \left(\frac{({\bf z}-{\bf s})^2}{8}-\phi \right)\right] ^{-(1+\mu)}
\end{align}
and we assume here $\mu>D/2$. The tail for large $s=|{\bf s}|$ is obtained, by manipulations
similar as the case $D=1$ as:
\bea
\label{asymptlarge2}
\rho(s) \simeq \frac{2^{2+\mu} \mu ~ s_x}{s^{2+2 \mu}} \text{  for large $s$}
\eea 
Interestingly going to higher dimensions allows the fluctuations of the particle motion to spread even more. To illustrate that fact one 
can compute the marginal shock density along ${\bf e}_x$ defined as:
\begin{align}
\rho (s_x) = \int _{s_{\bot}} \rho ({\bf s}) = s_x \theta _{s_x >0} \int _{s_{\bot}} D({\bf s})
\end{align}
After some integrations from Eq.(\ref{rhov2}) one finds:
\bea 
\rho(s_x) & = \mu^2 F_{\mu+1,D-1}^2 s_x \int_0^\infty dz  \int_{-\infty}^0 d\phi e^{ - F_{\mu,D} |\phi|^{\frac{D}{2}-\mu} } \nn \\
&\times \left[\left(\frac{(z+s_x)^2}{8}-\phi \right) \left(\frac{(z-s_x)^2}{8}-\phi \right)\right] ^{-(\frac{3-D}{2}+\mu)}
\eea 
hence a formula very similar to Eq.\ref{rhov}, but with a modified exponent $\tilde{\mu} = \mu -(D-1)/2$, leading to an 
asymptotic algebraic decay of the shock size along $x$ with exponent $\tau' = 2-D+2 \mu$. The 
thermodynamic condition $\mu > D/2$ again ensures that the normalization integral (\ref{normD}) exists.

\section{Elastic manifolds: recalling the general Flory argument} 
\label{floryargument}

We now check that the obtained values for the exponents agree with the general argument. 
For this we now recall the Flory argument given in [\onlinecite{biroli_top_2007}] for the directed polymer, which we
straightforwardly generalize to a manifold of internal dimension $d$ (internal coordinate $x \in R^d$) with $D$ 
displacement components $u \in R^D$. We consider that the random potential $V(x,u)$ lives in a total embedding
space dimension $d+D$ and has short range correlations with a heavy-tailed PDF (\ref{tail}) indexed by $\mu$. 
Assume that a piece of size $L$ (in $x$) explores typically $W \sim L^\zeta$ in dimension $D$.
The volume explored by the manifold is $L^d W^D$, hence the minimal value of $V$ on this volume
behaves as $\sim (L^d W^D)^{1/\mu}$. This leads to $\mu \theta = d + D \zeta$. Imposing again that
elasticity and disorder scale the same way (this is guaranteed by the general STS symmetry i.e. statistical invariance under
tilt) leads to $\theta = 2 \zeta+d - 2 $. Hence we obtain:
\bea
\zeta = \frac{d + \mu (2-d) }{2 \mu-D} \\
\theta= \frac{2 d + D (2-d) }{2 \mu - D } 
\eea 
with the (naive) threshold value beyond which one (presumably) recovers Gaussian disorder universality class:
\bea
\mu_c = \frac{d+D \zeta_{SR}}{d-2 + 2 \zeta_{SR}}
\eea 
where $\zeta_{SR}$ is the roughness exponent for SR Gaussian disorder. 
For $d=0$ one recovers the above values (\ref{exponentsD}) for the toy model in general
dimension $D$. For $d=1$ this gives the values given in [\onlinecite{biroli_top_2007}]
and recalled in the Introduction. It is interesting to note that at the upper-critical dimension
$d_{uc}=4$, $\zeta_{SR}=0$ hence the critical value is $\mu_c=2$.

\section{Conclusion}

In the present paper we have studied the toy model for the interface, i.e. a point in a random potential, in presence of
heavy tailed disorder with exponent $\mu$. In the scaling regime it leads to a universality class analogous to the Frechet class 
for extreme value statistics. 
It was found that all the relevant distributions (minimum energy, position, sizes of shocks) exhibit also power law tails with modified exponents
continuously dependent on $\mu$. Hence the presence of heavy-tails in the underlying disorder pervades through all observable and
modify the behavior for every value of $\mu$. That has to be compared with the directed polymer problem, where the effect of heavy tails disappears in favor of a "Gaussian" behaviour for $\mu>5$.  

In addition we have obtained here the shock size distribution for
an "exotic" example of
decaying Burgers turbulence, close from the Kida class because of the short range correlations in the initial potential, but
markedly different because of the heavy tails. 

Finally, because of these heavy tails the Functional RG method which, in its present form,
is based [\onlinecite{LeDoussal2008,LeDoussalWiese2008a}]
on the existence of the moments of the position of the minimum $u(r)$ cannot be applied in a standard way (at least in $d=0$). We hope our study will inspire progress on the more
general problem of the elastic manifold in the heavy tailed disorder. 
{\it Acknowledgements.} We thank J.P. Bouchaud for useful discussions.

\appendix

\section{Exotic regime in decaying Burgers turbulence} 
\label{app:Burgers} 

The above particle model is directly related to the Burgers equation for a velocity field ${\sf v}(r,t)$, 
a simplified version of Navier-Stokes used to model compressible fluids. 
\bea
\partial_t {\sf v}(r) = \nu \partial_r^2 - \frac{1}{2} \partial_r {\sf v}(r)^2 
\eea 
This equation can be integrated using the Cole-Hopf transformation.
Here we study only the inviscid limit (of zero viscosity $\nu=0^+$). In that case the solution is given by:
\bea
{\sf v}(r,t) = \partial_r H(r) = \frac{r - u(r)}{t}  
\eea 
in terms of (\ref{model1}) one defines the "time" $t$ as:
\bea
t = m^{-2}
\eea 
and the initial condition:
\bea
{\sf v}(r,t=0) = \partial_r H(r)|_{t=0}= \partial_r V(r) 
\eea 
where $V(u)$ is the bare disorder of the toy model. 
In this paper we focused on the case when $V(u)$
is short range correlated with a heavy tail. This corresponds
to a well defined but peculiar type of distribution for the initial velocity field: it also
has a tail exponent $\mu$, but exhibits local anti-correlations so that
$V(u)$ remains short range correlations (if ${\sf v}(r,t=0)$ was SR correlated 
with a heavy tail 

As is well known evolution from a smooth initial condition presents shocks in finite time,
i.e the velocity field ${\sf v}(r,t)$ does not remain continuous but presents (negative) jumps in a discrete set of locations $r_\alpha$ where ${\sf v}(r_\alpha^+,t)-{\sf v}(r_\alpha^-,t) = \Delta {\sf v} <0$. These corresponds to the (positive) jumps in $u(r)$, 
more precisely one has $\Delta {\sf v} = - S/t$ where $S$ is the dimension-full shock size $S=u_m s = m^{-\zeta} s$ with
the dimensionless size $s$ studied in the present paper. To translate our results in terms of
velocity jumps in Burgers, one thus just identifies $\Delta {\sf v} = - t^{\frac{\zeta}{2}-1} s$ (indeed the
length scale is $m^{-\zeta} = t^{\zeta/2}$), where $\zeta$ is given by 
(\ref{exponentsD}).

Finally the time dependence of the mean energy density $E$ is
given by $E = \frac{1}{2} {\sf v}^2 \sim t^{- (2-\zeta)}= t^{- 2 (\mu-D)/(2 \mu-D)}$, which recovers the result of [\onlinecite{gurbatov_universality_2000}]. Note that the regime $D/2 < \mu < D$ is very peculiar since 
it predicts an energy density growing instead of decaying, as discussed there.

\section{From infinite product to integral} 
\label{app:infinite} 

To understand better the convergence to the continuum limit let us first choose a
Pareto distribution, i.e. with a hard cutoff :
\bea
&& P_>(V) = \left(1-\frac{1}{(-V)^\mu} \right) \theta_{V<V_0} 
\eea 
and consider again the infinite product (\ref{infi_prod}). 
It can be rewritten, in the rescaled units i.e $u \to m^{-\zeta} u$, $V \to m^{-\theta} V$ as
(taking into account the Jacobian involved in the rescaling):
\bea \label{infi_prod2}
&& p(u,V)= m^{-(\zeta+\theta)} \frac{\mu}{(m^{-\theta} |V|)^{1+\mu} }  \theta_{V<V_0 m^\theta} \\
&& \times \prod _{u'\neq u} \theta(H-\frac{u'^2}{2}< V_0 m^{\theta}) e^{ \ln(1- m^{\mu \theta} (-H+\frac{u'^2}{2})^{-\mu}) } \nn
\eea
We see here that for $m \to 0$ it vanishes unless $H-\frac{u'^2}{2}<0$ for all $u' \neq u$, but since in that limit 
the lattice grid tends to continuum, this condition becomes equivalent to $H<0$. Since $V < H$ we do not need to
retain the constraint $V<0$. The infinite product becomes an
integral, the logarithm can be expanded, leading to:
\bea
&& p(u,V)= \frac{\mu}{|V|^{1+\mu} } \theta_{H<0} 
e^{- \int du'   (-H+\frac{u'^2}{2})^{-\mu}) } \nn
\eea
which leads to the result given in the text.

The mechanism holds for more general distributions with the same tail. As discussed in the 
text the rescaled $P_>(m^{-\theta} y)$ converges to unity for $y<0$ and to zero for 
$y>0$ so the precise shape of the distribution does not matter. More precisely, the weight of the
events with $H>0$ vanishes. To illustrate the point consider the worst case, i.e.
when $P_>(V)$ is slowly decaying on the positive $V$ side, e.g. as $V^{-\alpha}$. Then,
for $H>0$ (and $m \to 0$) there is an additional factor:
\bea
&& \approx \prod_{u'\neq u} \frac{\theta_{H - \frac{u'^2}{2} >0} }{m^{-\alpha \theta} (H - \frac{u'^2}{2} )^{\alpha} } \\
&& \simeq m^{\alpha \theta} e^{- \int_{-\sqrt{2H}}^{\sqrt{2H}} du' \ln(H - \frac{u'^2}{2})} = O( m^{\alpha \theta} )
\eea 
since the integral is convergent, and this factor kills the contribution of the events with $H>0$ (more precisely all the events with
$H > - m^{-\gamma}$ with any $0<\gamma<\theta$, in the original units). 

\section{Moments of $u$}
\label{app:moments} 

From (\ref{Pu}) and (\ref{psi}), we find the moments, for any real $n>0$ such that $2 n < 2 \mu-1$:
\bea
\overline{u^{2 n}} = F_\mu^{\frac{2 n}{2 \mu-1}} \frac{2^n \Gamma \left(n+\frac{1}{2}\right) 
\Gamma \left(\frac{\mu - \frac{1}{2} -  n}{\mu- \frac{1}{2} }\right) 
\Gamma \left(\mu+\frac{1}{2}- n \right)}{\sqrt{\pi } \Gamma \left(\mu +\frac{1}{2}\right)} \nn
\eea 
The $2n$-th moment thus diverges as $n \to \mu-\frac{1}{2}|^-$ as:

\bea
\overline{u^{2 n}} \simeq \frac{2^{\mu}}{\mu - \frac{1}{2} -  n} 
\eea

\section{Normalisation of the shock density}
\label{app:norma} 

A consistency check for the shock density is to check the normalisation given in Eq.\ref{norm1}, i.e. $\int ds s \rho(s)=1$. We recall that:
\begin{align}
\nonumber
I&= \int _{s>0} ds s \rho(s) = \frac{1}{2} \int_s s^2 D(s) =  \\
\nonumber
&=\frac{1}{2} \int du_1 du_2 d\phi ~ (u_1-u_2)^2  f\left(\phi-\frac{u_1^2}{2}\right) \\
&\times f\left(\phi-\frac{u_2^2}{2}\right)e^{- \int dz' F(\phi - \frac{(z')^2}{2})}
\end{align}
Due to the symmetry in the variables $(u_1,u_2)$, one can only consider for example:
\begin{align}
\nonumber
I_{u_1}&=\int  du_1 du_2 d\phi ~ u_1^2  f\left(\phi-\frac{u_1^2}{2}\right) \\
\nonumber
& \times f\left(\phi-\frac{u_2^2}{2}\right)e^{- \int dz' F(\phi -\frac{(z')^2}{2})}\\
\nonumber
=&-\int  du_1 d\phi u_1^2  f\left(\phi-\frac{u_1^2}{2}\right) \partial_{\phi}e^{- \int dz' F(\phi -\frac{(z')^2}{2})}\\
\label{int1}
=&\int  du_1 d\phi u_1^2  \partial_{\phi} f\left(\phi-\frac{u_1^2}{2}\right) e^{- \int dz' F(\phi -\frac{(z')^2}{2})}
\end{align}
where we used that, because of the limits $f(\phi) \to 0$ at $\phi \to -\infty$ and $F(\phi) \to \infty$ at $+\infty$, the boundary terms vanish. Considering the argument $\phi-u_1^2/2$ in $f(\cdot)$, one has the equivalence of the operators $\partial _{\phi} \leftrightarrow -u_1^{-1} \partial _{u_1}$ acting on $f(\cdot)$. Switching to $\partial_{u_1}$ derivatives in Eq. (\ref{int1}), and integrating by parts once again:
\begin{align}
\nonumber
I_{u_1}&=-\int  du_1 d\phi u_1  \partial_{u_1} f\left(\phi-\frac{u_1^2}{2}\right) e^{- \int dz' F(\phi -\frac{(z')^2}{2})}\\
\nonumber
&=\int  du_1 d\phi f\left(\phi-\frac{u_1^2}{2}\right) e^{- \int dz' F(\phi -\frac{(z')^2}{2})} \\
\nonumber
& =1
\end{align}
where again the boundary terms vanish due to $f(\phi - u^2/2) \to 0$ for $u \to \pm \infty$. Hence $I=\frac{1}{2}(I_{u_1}+I_{u_2})=1$ and the normalisation is properly recovered. The deeper reason behind these identities arises from the STS symmetry, i.e. the fact that the disorder
is statistically translationally invariant (see e.g. [\onlinecite{MonthusLeDoussal2004,LeDoussal2008}]).

Note that all the steps of this calculation easily generalize to higher $D$, the only change being that now
$u_1^2 \partial_\phi  \equiv - {\bf u}_1 \cdot \nabla_{{\bf u}_1} $ acting on $f(\phi - u_1^2/2)$. The final result
is then $I = D$ as discussed in the text. 

\section{The 2-points function}
\label{2pointfunction}

Let us consider the joint probability that $(V_1,u_1)$ and $(V_2,u_2)$ realize the minimum total energy respectively when the quadratic well is centered in $r_1$ and when it is centered in $r_2$, in the same realization of the disorder. The minimal energies are denoted by:
\bea
H_j = V_j + \frac{(u_j - r_j)^2}{2}  \quad , \quad j=1,2 
\eea 
This probability reads:
\begin{align}
&p(V_1,u_1,V_2,u_2) dV_1 du_1 dV_2 du_2 \\
&= f(V_1)f(V_2) dV_1 du_1 dV_2 du_2 \nn \\
\prod _{\substack{d V_j', du_j'\\u'_1 < u^*\\ u'_2 > u^*}} &\left( 1 - \theta_{V'_1 + \frac{(u'_1-r_1)^2}{2} < V_1 + \frac{(u_1-r_1)^2}{2}}  
 f(V'_1)dV'_1 du'_1 \right) \nn \\
&\times \left( 1 -  \theta_{V'_2 + \frac{(u'_2-r_2)^2}{2} < V_2 + \frac{(u_2-r_2)^2}{2}}  f(V'_2)dV'_2 du'_2 \right) \nn
\end{align}
where $u^*$ is the intersection abscissa of the two parabola, as represented in Fig.\ref{shockpic} given by:
\bea
H_1 - \frac{(u^* - r_1)^2}{2} = H_2 - \frac{(u^* - r_2)^2}{2}
\eea 
whose common value is denoted $\phi$ below. 
The additional Heaviside functions ensure that the random potential lies above these two parabola 
and touches those parabola on the two points $u_1$ and $u_2$.

The characteristic function can then be written:
\begin{align}
&\langle e^{\lambda (u(r_2)-u(r_1))}\rangle = \int dV_1 dV_2 du_1 du_2 
e^{\lambda(u_2-u_1)} \\
& \times \left(f(V_1) \delta_{V_2=V_1, u_2=u_1} +f(V_1) f(V_2) \theta_{u_1 < u^*<u_2} \right) \\
&\times e^{- \int _{u<u^*} F(H_1  - \frac{(u-r_1)^2}{2}) - \int _{u>u^*}F(H_2 - \frac{(u-r_2)^2}{2})}
\end{align}
where the first term accounts for the contribution when there is no shock between $r_1$ and $r_2$ and
the second when there is at least one. Let us now perform the change of variables:
\begin{align}
x&= \frac{r_2-r_1}{2} \text{ and } y=u^* - \frac{r_1+r_2}{2} \\
z&=u - r_1 \text{ and } z'= r_2 - u \\
z_1&=u_1 - r_1 \text{ and } z_2= r_2 - u_2\\
\phi &= H_1 -  \frac{(x+y)^2}{2} =  H_2 - \frac{ (x-y)^2}{2} 
\end{align}
hence $x+y=u^*-r_1$ and $x-y=r_2-u^*$. In terms of the
auxiliary functions:
\begin{align}
\nonumber
&J_+ (\phi , y,x) = \int_{z_1 \leq x+y } dz_1 f \left(\phi + \frac{(x+y)^2-z_1^2}{2} \right) e^{- \lambda z_1} \\
\nonumber
&J_- (\phi , y,x) = \int_{z_2 \leq x-y } dz_2 f \left(\phi + \frac{(x-y)^2-z_2^2}{2} \right) e^{- \lambda z_2} \\
\nonumber
&I_+(\phi , y,x) = \int_{z \leq x+y } dz F \left(\phi + \frac{(x+y)^2-z^2}{2} \right) \\
\nonumber
&I_-(\phi , y,x) = \int_{z' \leq x-y } dz' F \left(\phi + \frac{(x-y)^2-z'^2}{2} \right)
\end{align}
the characteristic function of the difference $u(r_2)-u(r_1)$ takes the form:
\begin{align} \label{2pt} 
&\langle e^{\lambda (u(x)-u(-x))}\rangle =  \\
&\int d\phi dy \left[ f(\phi) + 2x e^{2 \lambda x} J_{+}(\phi ,y,x)J_{-}(\phi ,y,x) \right] \nn \\
&\exp \left(-I_+ (\phi ,y,x)-I_- (\phi ,y,x) \right) \nn 
\end{align}
where the $2 x=r_2-r_1$ factor comes from the Jacobian $dV_1 dV_2 du_1du_2 = 2 x d\phi du^* dz_1 dz_2$. 

This formula generalizes to arbitrary $f(\phi)$ the one given in [\onlinecite{dbernard}] for a particular function $f(\phi)$. 
There it is given in terms of the (scaled) Burgers velocity field ${\sf v}(r) = r - u(r)$. One easily checks the
normalization i.e. that for $\lambda=0$ Eq. (\ref{2pt}) is a total derivative and integrates to unity. 

It is now rather straightforward to expand this formula to $O(x)$ and to recover the expression for the
shock density $\rho(s)$ given in the text using the identification (\ref{id1}).

\section{Asymptotics of the shock density}
\label{app:asympt}

The constant $C_\mu$ in the text can be obtained as:
\bea
&& C_\mu = \frac{\mu  (2 \mu -1) (2 \pi )^{\frac{\mu +1}{1-2 \mu }}}{3
   (4 \mu +1)} \\
&&   \times  \frac{
   \left(\frac{\Gamma \left(\mu -\frac{1}{2}\right)}{\Gamma (\mu
   )}\right)^{\frac{4 \mu +1}{1-2 \mu }} \Gamma \left(2 \mu
   +\frac{3}{2}\right) \Gamma \left(4+\frac{3}{2 \mu -1}\right)}{ \Gamma (2 \mu +2)}
\eea
where $C_\mu$ is an increasing function which vanishes at $\mu=1/2^+$ with an essential singularity
$C_\mu \simeq \exp \left( - \frac{3}{4} \frac{2-\ln(9/8)}{\mu-\frac{1}{2}} \right)$ and grows as
$C_\mu \simeq \frac{\mu^{3/2}}{2 \sqrt{\pi}}$ at large $\mu$.

\bibliographystyle{apsrev}
\bibliography{bibliokida,citation}

\end{document}